# An Integration Test Order Strategy to Consider Control Coupling

Shujuan Jiang, Miao Zhang, Yanmei Zhang, Rongcun Wang, Qiao Yu, Jacky Wai Keung

**Abstract**— Integration testing is a very important step in software testing. Existing methods evaluate the stubbing cost for class integration test orders by considering only the interclass direct relationships such as inheritance, aggregation, and association, but they omit the interclass indirect relationship caused by control coupling, which can also affect the test orders and the stubbing cost. In this paper, we introduce an integration test order strategy to consider control coupling. We advance the concept of transitive relationship to describe this kind of interclass dependency and propose a new measurement method to estimate the complexity of control coupling, which is the complexity of stubs created for a transitive relationship. We evaluate our integration test order strategy on 10 programs on various scales. The results show that considering the transitive relationship when generating class integration test orders can significantly reduce the stubbing cost for most programs and that our integration test order strategy obtains satisfactory results more quickly than other methods.

——————————  ◆  ——————————

## 1 INTRODUCTION

A common problem in the interclass integration testing of object-oriented software is to determine the order in which the classes are integrated and tested [1]. Theoretically, the function used by the first class to be tested must be mimicked by creating "stubs" (or mock objects). Suppose that a system has 32 classes and that 8 of those classes have been integrated. Among the remaining 24, suppose that class A depends on class B. If class A is integrated before class B, a test stub that simulates the behavior of B must be generated for A to use until B becomes available. That is, a stub (or a mock object) is an incomplete class that provides the signatures necessary for full compilation and integration, but does not implement the full functionality. Although testing and mocking frameworks (e.g., JUnit 5 [2] and Mockito [3] for Java) have been developed to facilitate the simulation of class dependencies, subbing[1] is still an expensive and error-prone operation. This problem is generally called the class integration and test order (CITO) problem. Different test orders often lead to different testing cost; a proper test order can significantly reduce the testing cost. Therefore, the accurate determination of an optimal test order is an important problem. In other word, an optimal class test order for an object-oriented program is important because it can save redundant efforts in constructing stubs for integration testing. Several publications have proposed various approaches and strategies for the CITO problems and the evaluation metrics for the results obtained with their methods.

The stubbing cost is used to measure various CITOs. Studies have assessed the stubbing cost from several aspects. Kung et al. [1] first used the number of stubs to evaluate the performance of approaches. Fewer stubs were needed, and the method's performance was better. This metric was once popular among researchers, but later studies found that the class test order with the fewest stubs may not be the best choice. The metric that counts only the number of stubs omits other factors that may affect the stubbing cost. Thus, Briand et al. [4] suggested a metric of stubbing complexity to measure the stubbing cost. They adopted attribute complexity and method complexity to record the number of method invocation and number of attribute access among classes and advanced a linear-weighted equation to calculate the stubbing complexity. The minimum value of the stubbing complexity indicates the least effort in creating stubs. On this basis Abdurazik et al. [5] identified four kinds of information on coupling measures (including the number of parameters and the number of return value types) and improved the previous method for measuring the complexity of stubs.

The aforementioned information on coupling measures in stubbing complexity mainly involved interclass data coupling, such as the number of attributes accessed and number of methods invoked, but it omitted control coupling, which is also common in interclass relationships. For example, Kung et al. [6] noted that "the inheritance and aggregation relationships involve not only control coupling, but also data coupling." To avoid constructing error-prone stubs and to minimize the testing effort, Briand et al. [7] believed that "stubs should remain as simple as possible, and ideally only contain sequential control flow". Control dependences between two statements may cause interclass control coupling, and the latter can form interclass indirect relationships that should be a crucial factor in measuring the stubbing cost.

• Shujuan Jiang, Yanmei Zhang, and Rongcun Wang are with the Mine Digitization Engineering Research Center of the Ministry of Education, School of Computer Science and Technology, China University of Mining and Technology, Xuzhou 221116, China (e-mail: shjjiang@cumt.edu.cn, ymzhang@cumt.edu.cn, rcwang@cumt.edu.cn). Miao Zhang and Jacky Wai Keung is with the department of Computer Science, City University of Hong Kong, Kowloon, Hong Kong (e-mail: miazhang9-c@my.cityu.edu.hk, Jacky.Keung@cityu.edu.hk). Qiao Yu is with school of Computer Science and technology, Jiangsu Normal University, Xuzhou 221116, China (e-mail: yuqiao@jsnu.edu.cn).

_____________________

[1] In the generation of class integration test orders, we focus only on which areas are emulated in class dependencies, rather than how they are emulated.

In this paper, we do not distinguish mocks and stubs, and we use the term "stubbing" to refer to the creation of a stub or mock object.

This means that control coupling should also be considered when referring to the stubbing complexity. However, nearly all existing methods evaluate the stubbing complexity from only four aspects: attribute access, method invocation, types of return, and parameters. Control coupling between classes has rarely been studied in CITO problems.

To address this issue, we propose an integration test order strategy to consider control coupling called ConCITO (**Con**trol Coupling **C**lass **I**ntegration **T**est **O**rder Generation). ConCITO improves the existing stubbing cost measurement methods by adding a new kind of coupling measure information (i.e., control coupling) and designs a new formula of stubbing complexity that combines data coupling with control coupling. Specifically, ConCITO first performs program analysis to identify interclass indirect relationships and designs an extended object relation diagram (EORD) to describe this kind of relationship. Second, ConCITO devises a new method of calculating the complexity of stubs created for each relationship. Unlike existing methods of stubbing complexity measurement that consider only attribute complexity and method complexity, the new stubbing complexity includes control complexity, which measures the control coupling based on its probability of occurrence. The statements that form each inter-class indirect relationship, the control dependences of these statements, and the paths that contain these dependences can be identified, and the probability of the path being executed in the control flow graph can be calculated. Finally, ConCITO adopts the new assessment method of stubbing complexity to generate CITOs.

We evaluated ConCITO by conducting experiments on 10 programs. The results show that control coupling could indeed affect the stubbing cost of a CITO; ConCITO more accurately assessed the performance of class test orders and thereby generated the class test order with the minimum stubbing cost. The contributions of this paper are as follows.

An approach is proposed to describe the interclass indirect relationships caused by control coupling. To the best of our knowledge, this kind of indirect relationship has never been leveraged as coupling information.

An EORD is constructed to describe interclass relationships more comprehensively, including the familiar relationships such as inheritance and association, and the newly proposed indirect relationships.

A measurement method is devised to calculate stubbing complexity from the aspects of both data coupling and control coupling. For each interclass indirect relationship, an equation is derived to calculate the coupling complexity based on the possibility that control coupling occurs.

The remainder of this paper is organized as follows. Section 2 introduces the motivation; we present our approach and its architecture in Section 3. The experiments follow in Section 4, related work is presented in Section 5, and Section 6 concludes this report.

## 2   MOTIVATION

In this section, we explain the reasons for interclass indirect relationships and illustrate how these indirect relationships affect the stubbing cost of CITOs with a sample program (Fig. 1).

The sample program contains three classes: A, B, and C. Class A invokes methods B1 and B3 in class B through methods A1 and A3 (lines 8, 12, and 20). Similarly, class B calls methods C1 and C2 in class C through methods B1 and B3 (lines 6 and 12). In fact, class A might invoke methods in class C by calling the abovementioned methods in class B, and classes A and C can thus form an interclass indirect relationship.

| | | | |
|---|---|---|---|
| 1 | public **class A**{ | 15 | void **methodA2**(){ |
| 2 | int a, b = 1; | 16 | …. |
| 3 | boolean is; | 17 | } |
| 4 | B b = new B(); | 18 | int **methodA3**(int x, int y){ |
| 5 | void **methodA1**(double m, int n){ | 19 | if(x>3 \|\| y < 5) |
| 6 | if(m < 1.2){ | 20 | return b.methodB3(b) + 1; |
| 7 | if(n < 5) | 21 | else… |
| 8 | b.methodB1(a); | 22 | return 0; |
| 9 | else a = 3; | 23 | } |
| 10 | } | 24 | void **methodA4**(){ |
| 11 | if(is == true){ | 25 | …. |
| 12 | b.methodB1(a); | 26 | } |
| 13 | } | 27 | } |
| 14 | } | | |
| 1 | public **class B**{ | 8 | void **methodB2**(){ |
| 2 | C c = new C(); | 9 | a.methodA2(); |
| 3 | A a = new A(); | 10 | } |
| 4 | void **methodB1**(int x){ | 11 | int **methodB3**(int x){ |
| 5 | if(x > 2) | 12 | int t = c.methodC2() * 3; |
| 6 | c.methodC1(); | 13 | if(x > 0) |
| 7 | } | 14 | … |
| | | 15 | return t; |
| | | 16 | } |
| | | 17 | } |
| 1 | public **class C**{ | 7 | int **methodC2**(){ |
| 2 | int c = 5; | 8 | ….return c; |
| 3 | A a = new A(); | 9 | } |
| 4 | void **methodC1**(){ | 10 | void **methodC3**(){ |
| 5 | …. | 11 | a.methodA4(); |
| 6 | } | 12 | } |
| | | 13 | } |

Fig. 1. Sample program

The method invocation among the three classes is shown in Fig. 2(a). Fig. 2(b) shows the object relation diagram (ORD) constructed with the existing methods, which do not involve the indirect relationship between classes A and C. Assuming that we adopt a graph-based algorithm to generate CITOs with the minimum number of stubs, CITOs can be generated by removing edges and breaking cycles in the ORD. The removal of fewer edges involved in more cycles means that fewer stubs need to be created, so we remove the edge "A→B" to break both cycles and generate the class test order "A→C→B". Class A calls methods of class B, but class B cannot provide reliable services because it has not been tested, so we create stubs for class A to emulate the behaviors of class B. However, the result has two limitations. First, the class test order is suboptimal because it does not break cycles between classes A and C. Indeed, three classes form three cycles when considering the indirect relationship between classes A and C (shown as Fig. 2(c)). Second, when constructing stubs to provide class B's services for class A, we should consider the possible effects of class $C$ on these stubs because some methods (i.e., methods $C1$ and $C2$) in class $C$ that are invoked by class $B$ may affect the behavior of those methods in class $B$ that are invoked by class $A$. This kind of stub should be avoided because it is more error-prone and expensive. Omitting interclass indirect relationships could influence the CITOs generated by the existing algorithms and the number of stubs required.

The example in Fig. 1 illustrates how control coupling affects stubbing. Table 1 shows a comparison of constructed stubs under two situations, that is considering control coupling or not. Columns 2 to 6 show the cycles, deleted edges, the classes that are required to be stubbed, the methods that are required to be stubbed, and the corresponding test order, respectively. Where "1" means not considering the indirect dependencies and "2" means considering the indirect dependencies. We can see that class $B$ is the stubbed class when we do not consider the indirect dependencies, and class $A$ (or classes $A$ and C) is the stubbed classes when we consider the indirect dependencies. Therefore, a stub is a (or some) simulated called method(s) or function module such as method $B1$ or $B3$ in the example of Fig. 1.

To test the interactions between classes $A$ and $B$, specifically, methods $A3$ and $B3$, we require a stub to emulate the return value of method $B3$ (line 20). If we do not consider the third method $C2$, we might return any integer values because method $B3$ does not restrict it. However, when we consider the indirect relationship between classes $A$ and $C$, we find that the return value of method $B3$ should be 15 (5 × 3) rather than a random value. Moreover, if the invocation statement of $b.methodB3$ is included in a condition statement, such as "$if$" or "$while$", the execution trace of program would be significantly different largely due to different return values of method $B3$. For example, if there is a code snippet as follows:

```
if (b.methodB3 == 15)
    // statement1;
else
    // statement2;
```

It is obvious that $statement1$ should be executed if we return the correct value, but if we do not consider class $C$, it is likely to return a random value and then go to $statement2$, which is against what was intended in the program. Therefore, consideration of control coupling can create a more accurate stub.

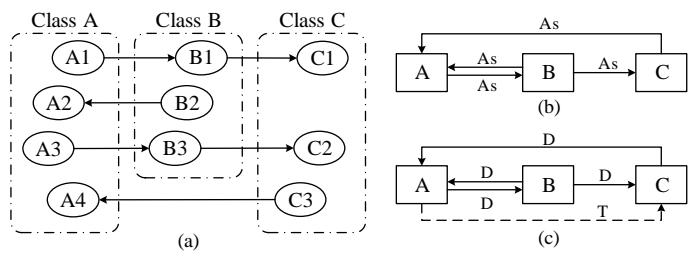

Fig. 2. ORD of sample programs

TABLE 1
COMPARISON OF CONSTRUCTED STUBS UNDER TWO SITUATIONS

|  | Cycles | Deleted edges | Stubbed classes | Stubs | Test oder |
|---|---|---|---|---|---|
| 1 | A→B→A<br>A→B→C→A | A→B | B | B1 or B3 | A->C->B |
| 2 | A→B→A<br>A→B→C→A<br>A→C→A | B→A&<br>C→A<br>or<br>B→A&<br>A→C | A<br>or<br>A&C | A2 or A4<br>or<br>A2&A4<br>or<br>A2&C1<br>or<br>A2&C2<br>or<br>A2&C1&<br>C2 | C->B->A |

Our experimental results in Section 4.4 confirm the existence of this kind of indirect dependencies.

For other kinds of approaches, such as search-based algorithms that aim at minimizing stubbing complexity, the overall stubbing complexity is used to direct the evolution of individuals in a population. The two kinds of overall stubbing complexity that are calculated by considering control coupling or not might lead to different directions for evolution, so the CITOs generated in different directions and their stubbing costs may be distinct. Omitting control coupling in interclass indirect relationships results in an inaccurate stubbing complexity measurement, and the inaccurate overall stubbing complexity calculated by this kind of measurement may cause deviation of guidance that lead to failure of the search process.

Because interclass indirect relationships affect the stubbing cost of CITOs and thereby influence the performance of various approaches, this study identifies and analyzes the interclass indirect relationships in object-oriented programs and proposes a new measurement that considers control coupling to calculate stubbing complexity. The details of our work are introduced in Section 3.

## 3 APPROACH

To tackle the problem in which the existing measurement methods of stubbing cost lack the assessment of control coupling, our approach ConCITO first constructs an extended object relation diagram (EORD), extracts the control coupling among the interclass indirect relationships and measures their control complexity, and finally designs a new measurement for stubbing complexity. The procedure of ConCITO, shown as Fig. 3, consists of three steps: preparation, calculation of the stubbing complexity, and generation of CITOs.

Step 1: Preparation. We first give the definition of the inter-class indirect relationships like the relationship between classes $A$ and $C$ shown in the sample program (Fig. 2(c)). Then, according to the definition, we execute static analysis on programs, find all relationships among classes, and construct an EORD that is extended on the basis of the previous ORD by adding the above indirect relationships. Noting that the relationship between two classes may contain both direct relationships and indirect relationships, and different kinds of relationships result in different coupling measures information, we identify these kinds of relationships from EORD in preparation to extract three kinds of coupling measures: attribute coupling, method coupling, and control coupling.

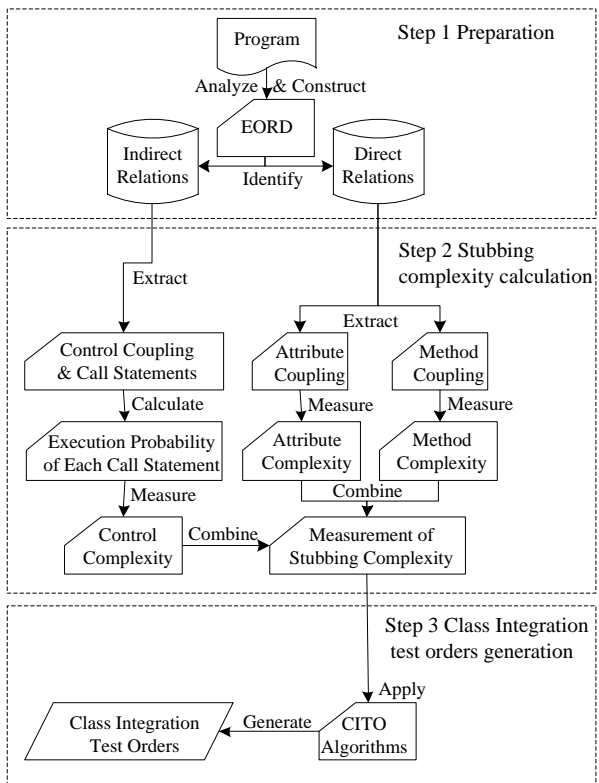

Fig. 3. Procedure of ConCITO

Step 2: Stubbing complexity calculation. To devise a new measurement of stubbing complexity, we first calculate the complexities of attribute coupling, method coupling, and control coupling, respectively. As with previous methods, the number of interclass attribute access can be counted as the attribute complexity between the two classes, and similarly, the number of interclass methods invoked can be calculated as the method complexity. As for the control complexity, we first identify control coupling according to the indirect relationship between two classes and obtain all intermediate classes involved in indirect relationships. We can then extract call statements that connect two contiguous classes and generate path conditions for these call statements. Next, the execution probability of each call statement can be calculated based on its path conditions. Because the control coupling between two classes exists only when all call statements that connect two contiguous classes are actually executed, the control complexity can be measured by the possibility of all call statements being executed. Finally, measurement of the stubbing complexity can be accomplished by combining the attribute complexity, method complexity, and control complexity.

Step 3: Generation of CITOs. The new measurement of stubbing complexity can be applied in CITO algorithms,

such as the existing graph-based methods or search-based methods, to generate the class test orders with the minimum stubbing complexity. The details of the three steps are described below.

## 3.1 Extended Object Relation Diagram

This section presents an EORD that describes the relationships among classes in an object-oriented program at full length. The EORD is constructed based on the traditional ORD by adding the new indirect relationship, that is, the transitive relationship. For example, the relationship between classes $A$ and $C$ is shown in the sample program (Fig. 2(c)).

**Definition 1. Transitive relationship (TR)**. In an object-oriented program $P$, when class $i$ depends on class $k$ and class $k$ is dependent on class $j$, the relationship between class $i$ and class $j$ can be established through class $k$. For example, class $i$ accesses $k.variable$ (an attribute in class $k$), and $k.variable$ can be obtained by a method in class $j$ (e.g., a method's return value of class $j$ is assigned to $k.variable$); or class $i$ invokes $k.method$ (a method in class $k$) and $k.method$ invokes another method or accesses another attribute in class $j$. This kind of indirect relationship between classes $i$ and $j$ can be defined as a transitive relationship.

As shown in the sample program in Fig. 2(a), class $A$ takes class $B$ as a medium and invokes methods $C1$ and $C2$ in class $C$, forming the transitive relationship. In fact, the relationships among classes are more complex in actual applications, so this kind of transitive relationship might involve not only three classes, but form a chain that connects multiple classes.

**Definition 2. Transitive relationship chain (TRC)**. Let $V$ denote the set of classes for the object-oriented program $P$; a transitive relationship chain from class $i$ to class $j$ ($i, j \in V \wedge i \neq j$) is defined as an execution path from a method or an attribute $x$ in class $i$ to a method or an attribute $w$ in class $j$, i.e., $x \rightarrow y \rightarrow \ldots \rightarrow w$. The remaining nodes in the path, such as node $y$, represent the method or attribute in other class $k$ ($k \in V \wedge k \neq i, j$). The execution path should satisfy the following conditions: (1) an attribute access or a method invocation must appear in two contiguous classes that are represented by adjacent nodes in the execution path. For instance, class $i$ should contain at least one method $x$ that invokes a method $y$ or at least one attribute $x$ that can be obtained by a method $y$ in class $k$. (2) All nodes in the directed path can appear only once, that is, no methods or attributes can be executed repeatedly.

**Definition 3. Extended object relation diagram**. An EORD for the object-oriented program $P$ can be represented by a direct diagram $EORD = (V, L, E)$. $V$ denotes the set of classes for the program $P$; $L = \{D, T, C\}$ denotes the set of types for the interclass relationships; and $D, T$, and $C$ are labels for direct relationship, transitive relationship, and combination relationship, respectively. $E = E_D \cup E_T \cup E_C$ denotes the set of interclass relationships, and $E_D$ contains only direct relationships, such as inheritance, aggregation, and association. $E_T$ represents the transitive relationship mentioned in definition 1, and $E_C$ denotes the relationships between two classes that contain both direct and indirect relationships.

According to these definitions, we can describe the inter-class relationships in more detail. Fig. 2(a) shows two transitive relationship chains from class $A$ to class $C$ according to the definition 2, that is, method $A1\rightarrow$method $B1\rightarrow$method $C1$, and method $A3\rightarrow$method $B3\rightarrow$method $C2$. Fig. 2(c) shows the EORD constructed for the sample program according to definition 3. Based on the EORD, we can extract the control coupling from the transitive relationship and the combination relationship to calculate the complexity of control coupling.

**Definition 4. Control coupling.** The transitive relationship between classes $i$ and $j$ (class $i$ depends on class $j$) means that whether or not some methods in class $i$ can be invoked or some variables in class $i$ assigned, are subject to class $j$. Control coupling is used to describe this kind of constraint involved in a transitive relationship, and this coupling includes two types: (1) some other classes' methods that are invoked by class $i$ require the services supplied by class $j$ (e.g., the attributes of class $j$ as the abovementioned methods' parameters); (2) some other classes' attributes that are accessed by class $i$ should be obtained by class $j$ (e.g., the return values of the methods in class $j$). The control coupling can be extracted from the relationships according to definition 4 and used to estimate the control complexity.

## 3.2 Control Complexity

The transitive relationship between classes $i$ and $j$ can be established only when at least one transitive relationship chain contains these two classes. Whether the program can be executed in accordance with the transitive relationship chain is restricted to the test cases in the integration testing and can be acknowledged only when the program is actually running. However, the probability of each transitive relationship chain being executed can be deduced by static analysis. The probability that a transitive relationship chain is executed would be uncertain, because the execution possibility differs for each call statement that connects two contiguous classes in the chain, and thus the transitive relationship might not really exist. To estimate the complexity of stub construction for a transitive relationship, we propose a measurement method for control complexity based on the statements' execution probability. The control complexity can be measured by the possibility of the related transitive relationship being established and is given by equation (1):

$$T(i, j) = 1 - \prod_{k=1}^{N} (1 - t_{ij}(k)) \qquad (1)$$

Let $T(i, j)$ denote the control complexity between classes $i$ and $j$, measuring the possibility that the transitive relationship between classes $i$ and $j$ really exists. Supposing that $N$ transitive relationship chains exist between classes $i$ and $j$, the transitive relationship breaks down only when none of these chains are executed. $t_{ij}(k)$ represents the possibility of the $k$-th transitive relationship chain being executed and can be calculated as equation (2):

$$t_{ij}(k) = \prod_{x=1}^{M} pc(x), x \in TRC \qquad (2)$$

Where a set $TRC$ contains all call operations (such as method invocation and attribute access) involved in the $k$-th transitive relationship chain. $pc(x)$ denotes the possibility that the $x$-th call operation is being executed; that is, the probability of the statements that contain the $x$-th method invocation or attribute access being run. Assuming that the number of call operations in $TRC$ is $M$ (including both method invocation and attribute access), execution of the transitive relationship chain is dependent on every call operation, so the possibility of the transitive relationship chain being executed (i.e., $t_{ij}(k)$) is equal to the product of all probabilities that each call operation is running (i.e., $pc(x)$).

To calculate the possibility that the $x$-th call operation is being executed ($pc(x)$), we should first locate all statements that contain the $x$-th call operation and extract the control dependences for each statement. We can then obtain the path conditions from the control dependences and calculate the execution possibility of each statement. Whether a statement can be executed is subject to the related branch statements, such as condition statements like "*if*" or "*switch*" and loop statements like "*while*" and "*for*" The statement can be executed only when its path conditions are satisfied. The program might contain more than one statement to execute the $x$-th call operation; thus $pc(x)$ is equal to the sum of all related statements' execution possibility. The calculation process of $pc(x)$ includes three steps: (1) extraction of statements and generation of path conditions; (2) calculation of each statement execution probability; (3) calculation of the call operation execution probability, which is introduced in detail in the next section.

### 3.2.1 Statements Extraction and Path Condition Generation

Fig. 4 shows the algorithm of statement extraction and path condition generation. Given an operation $x$ in class $C$, this algorithm first locates all statements that execute the operation $x$ and records them in the set, $StmtSet$. The algorithm then generates path conditions for each statement and records the mapping between a statement and its path conditions in the form of "key-value" in the set $PcMap$.

The algorithm begins with the initialization procedure (lines 1 and 2) where two output sets $StmtSet$ and $PcMap$ are initialized and set to an empty set. Every statement in class $C$ would be traversed to identify whether it contains the operation $x$ (lines 3 and 4): if it does, the statement would be added into $StmtSet$ (line 5), and the function for path condition generation would be invoked (line 6). The detail of the function for path condition generation is introduced in lines 9 through 20. For each statement, the control flow graph for the method that contains the statement is constructed (line 10). The algorithm then obtains the node that represents the basic block that contains the statement in the graph (line 11), traverses all parent nodes of the current node in the opposite direction of the edge from bottom to top until no parent node remains, and obtains an access path from the parent node of the original node to the entry node (lines 12 through 14). In the process of recursive visiting, if a node represents the basic block that contains branch statements (line 15), the predicate in the branch statement is extracted as a path condition for the node (line 16). All path conditions could be generated after all nodes in the access path are visited to identify whether they contain branch statements, and these path conditions would be saved in the set $PcMap$ (line 19).

**Algorithm:** Statements extraction and path condition generation
**Input:** the operation $x$ in class $C$
**Output:** a set of statements *StmtSet*; a set of path condition *PcMap*;
**begin**

1.    StmtSet←Φ
2.    PcMap←Φ
3.    **for** each statement *stmt* in class $C$ **do**
4.      **if** *stmt*.contains($x$) **then**
5.        StmtSet.add(*stmt*)
6.        genPathCondition(*stmt*)
7.      **end if**
8.    **end for**

9.    **function** genPathCondition(*stmt*)
10.   Control_Flow_Graph = genCFG()
11.   node←Control_Flow_Graph.getBasicBlock(*stmt*)
12.   temp←node
13.   **for** temp.getParentNode() != null **do**
14.     temp←temp.getParentNode()
15.     **if** temp.containsBranchStmt() == true **then**
16.       path_condition.add(temp.getCondition())
17.     **end if**
18.   **end for**
19.   PcMap.put(*stmt*, *path_condition*)
20.  **end**
**end**

Fig. 4. Algorithm of statements extraction and path condition generation

### 3.2.2 Statement Execution Probability

After the first step, we can obtain all statements that execute the $x$-th call operation in the transitive relationship chain and their path conditions. The execution probability can be calculated for each call statement according to the possibility that its path conditions will be satisfied. When the path condition is null, the statement can be executed sequentially and would not be interrupted by the jump statements, so its execution probability is one. The path conditions for multiple call statements are identical, which indicates that these statements are all in one basic block; that is, if one statement is run, the other statements would also be executed, so these statements all have the same execution probability.

The path conditions for other call statements are extracted from branch statements, and the various kinds of branch statements determine various execution probabilities. Hence, we discuss the execution probability of call statements based on their type. For conditional statements, such as "*if*" statements, the path conditions are a few inequalities. If there is only one inequality, the path condition is a simple path condition that does not contain an AND relationship and an OR relationship, and the execution probability of the call statement can be considered as 0.5, presumably in the absence of other information for reference. In contrast to the single inequality, when there are multiple inequalities and some may be contradictory (for instance, $x > 1$ and $x < 0$), the path conditions are in conflict, and the corresponding call statement cannot be executed. Conversely, if the path condition is a compound path condition that does not contain paradoxical inequalities, supposing that the number of inequalities is $N$ and that all of these inequalities are connected with an AND relationship, the execution probability of the call statement can be calculated as $1/2^N$, presumably on the condition that some information is lacking to confirm whether the inequality is true. For example, when the path condition is $x > 2$ && $y < 7$, there are four combinations of inputs for $x$ and $y$, and only one kind of input can satisfy the path condition; thus, the possibility that a combination of inputs satisfies the path condition is 0.25. Similarly, assuming that these inequalities are all connected with an OR relationship, the execution probability of the call statement can be calculated as $(1-1/2^N)$, presumably if we do not have enough information to verify that each inequality is true or false. For instance, if the path condition is $x < 3$ || $y > 6$, the number of combinations of inputs for $x$ and $y$ is also 4, and only one kind of input cannot satisfy the path condition, so the probability that a random kind of combination of inputs can meet the path condition is 0.75. For the other conditional statements, such as "*switch*" statements, the principle of calculating the execution probability of call statements is basically identical to that stated above. Supposing that the number of code blocks related to "*case*" and "*default*" statements is $N$, the probability that the call statement in a code block can be executed is $1/N$ if some reliable information for reference is lacking.

For looping statements such as "*for*" and "*while*" the call statements in the code block of the looping indicate that software developers expect that these statements can be executed repeatedly; thus, the execution probability of this kind of call statements is 1.

The prerequisite of the above discussion is that the execution of each statement is independent of the others; that is, the one call statement for the $x$-th call operation is not in the branch statement as the other call statement's path condition. However, in fact, this kind of situation might occur. If the call statement is one of its own path conditions, we first calculate the execution possibility for the call statement as a path condition and then calculate the execution probability for the latter according to the probability of the former.

### 3.2.3 Call Operation Execution Probability

According to the results obtained in the second step, we can calculate the execution probability for the $x$-th call operation in the transitive relationship chain. For the $x$-th call operation, it can be executed by multiple call statements, and each statement is independent. The $x$-th call operation cannot be run only if none of these statements are executed. Thus, the execution probability of the $x$-th call operation (i.e., $pc(x)$) is subject to the probability that each call statement can be executed

The execution probability of call operation can be calculated as the above three steps and used to calculate the control complexity for the transitive relationship. In what follows, we use the sample program in Fig. 1 as an example to illustrate how to calculate the control complexity between classes $A$ and $C$ (i.e., $T(A, C)$). It is easy to see that there are two transitive relationship chains from class $A$ to class $C$ in Fig. 2(a). The first step is to calculate the execution possibility for each transitive relationship chain. The first transitive relationship chain, that is, method $A1$→method $B1$→method $C1$, includes two call operations. The first call operation, method $A1$ invoking method $B1$, is implemented in class $A$ at lines 8 and 12. The path condition for the former statement at line 8 is $\{m < 1.2\ \&\&\ n < 5\}$, and both inequalities should be true, so the execution possibility for the statement is 0.25 ($0.5*0.5=0.25$). Similarly, the path condition for the latter statement at line 12 is a single equality, $\{is == true\}$, and the execution possibility for the latter is 0.5. Method $B1$ cannot be invoked only when both two statements cannot be run; hence, the possibility that method $B1$ can be invoked is 0.625 ($1-0.75*0.5=0.625$). The second call operation, method $B1$ invoking method $C1$, is implemented in class $B$ at line 6, and its execution possibility is 0.5. Thus, the execution possibility for the first transitive relationship chain is equal to the product of the probabilities that the above two call operations are run (0.3125). We can calculate the execution possibility for the second transitive relationship chain; that is, method $A3$ →method $B3$→method $C2$, in the same way. The former call operation, method $A3$ invoking method $B3$, is implemented in class $A$ at line 20 with the path condition $\{x > 3\ ||\ y < 5\}$; the latter call operation, method $B3$ invoking method $C2$, is implemented in class $B$ at line 12 with the path condition being null. The execution possibilities for the former and the latter are 0.75 ($1-0.5*0.5=0.75$) and 1, respectively. Thus, the value of the execution possibility for the second transitive relationship chain is 0.75. If the program would not be executed in accordance with one of the two transitive relationship chains, the transitive relationship between classes $A$ and $C$ would not be established, and thus the value of the control complexity between them would be zero. Therefore, according to the principle mentioned above, the value of the control complexity between classes $A$ and $C$ is 0.8281. The process of calculation for $T(A, C)$ is shown in Fig. 5.

Fig. 5. Calculation of T(A, C)

### 3.3 Stubbing Complexity

In the integration test, when class $i$ is dependent on class $j$ and the test order of class $i$ precedes that of class $j$, a stub that can be denoted as $stub(i, j)$ would be created for class $i$ to emulate the behavior of class $j$. The stubbing complexity that can be denoted as $SCplx(i, j)$ is used to measure the cost to construct $stub(i, j)$. The previous measurement methods for stubbing complexity evaluate only the attribute coupling and method coupling and omit the indirect call operations (such as method invocation and attribute access) via intermediate classes. Therefore, control complexity is proposed to measure the cost of stubs that are constructed for the interclass transitive relationship, and the existing formula of stubbing complexity has been improved with the addition of the control complexity. The improved formula, $SCplx(i, j)$ can be calculated with equation (3):

$$SCplx(i, j) = \left[ W_A \cdot \overline{A(i, j)}^2 + W_M \cdot \overline{M(i, j)}^2 + W_T \cdot T(i, j)^2 \right]^{\frac{1}{2}} \quad (3)$$

Where $A(i, j)$, $M(i, j)$, and $T(i, j)$ represent the complexity of attribute coupling, method coupling, and control coupling, respectively. The complexity of attribute coupling and method coupling are equal to the number of attributes and the number of methods emulated in $stub(i, j)$ respectively. Considering that the effect of control coupling might be masked when the absolute values of the complexities of attribute coupling and method coupling are much greater than those of control coupling, the complexity of attribute coupling and method coupling should be normalized by dividing by the maximum attribute coupling in the system. The values of $T(i, j)$, normalized $A(i, j)$, and $M(i, j)$ are all within [0, 1]. To ensure the same effect of the three factors, three nonnegative weights—$W_A$, $W_M$, and $W_T$ ($W_A+W_M+W_T=1$)—are all set to 1/3.

In an EORD, interclass relationships include three types: direct relationships, transitive relationships, and combination relationships (Section 3.1). The stubbing complexity for direct relationships such as inheritance, aggregation, and association contains only the first two items, and the value of the complexity of control coupling is zero. In contrast, the stubbing complexity for transitive relationships involves only control coupling. For the combination relationship, because the relationships between two classes contain both direct and indirect relationships, the stubbing complexity includes all three factors.

The overall stubbing complexity measures the stubbing cost in the integration test. Supposing that classes in the program will be integrated and tested according to the test order $o$ and that a set $Stubs$ contains all stubs required by the classes. The overall stubbing complexity, $OCplx(o)$ is equal to the sum of the complexity of stubs, $SCplx(i, j)$, which can be estimated with equation (4):

$$OCplx(o) = \sum_{(i, j) \in Stubs} SCplx(i, j) \quad (4)$$

where the stubbing complexity of $stub(i, j)$ can be calculated with equation (3).

### 3.4 CITO Algorithm

The improved formula for stubbing complexity can be adopted for various kinds of existing methods, such as graph-based algorithms and search-based algorithms.

For graph-based approaches, we first construct an EORD for the program and assign the weight for each edge in the diagram based on the stubbing complexity that considers the measurement of control coupling. Tarjan's algorithm [8] can then be applied to identify strongly connected components (SCCs), and Depth First Search algorithm would be used to find all cycles in each nontrivial SCC. We calculate the cycle-weight ratio for each edge that represents the weakly connected relationship in cycles, that is, the ratio of the number of cycles in which the edge is involved to the weight of the edge. To break more cycles with a lower stubbing cost, the edge with the highest cycle-weight ratio would be removed, and consequently, the number of cycles where each edge is involved should be recalculated. Edges are removed according to the principle described above until no cycles exist in the diagram. Finally, we can generate the CITOs by topological sorting for the acyclic diagraph.

For search-based methods, we first initialize test orders and then adopt evolution operators to produce new offspring. The improved formula for stubbing complexity is used as the fitness function to guide the evolution of populations and to select the optimal test order with the minimum stubbing cost.

We wish to apply this new measurement method for stubbing complexity to our previous work [9], and leverage an incremental strategy in which classes are integrated based on their test priority. In this strategy, we propose the concept of "testing cost" to measure the cost caused by constructing stubs that are required by a class and the concept of "test profit" to evaluate the convenience provided by the class for the process of testing subsequent classes. The testing cost and test profit are calculated based on the stubbing complexity between two classes using the improved equation (3). By comparing the test profit and testing cost, the test priority for each class can be adjusted using a reward and punishment mechanism; that is, the test priority for the classes with a positive test profit can be boosted, and that for the classes with a negative test profit should decrease. Each time after adjusting test priority, the classes with zero testing cost can be added into the class test order because they do not require stubs; otherwise, the classes with the highest test priority are added into the class test order. For other unintegrated classes, a multilevel feedback strategy can be used to recalculate their testing cost and test profit. If the unintegrated class is dependent on a tested class, its test profit would increase due to the reduction in its testing cost; in contrast, if the unintegrated class is dependent on a tested class, its test profit decreases. The above process continues until no classes remain. The CITO-generation algorithm in ConCITO can be outlined as follows:

Step 1. For any two classes in a program, calculate the stubbing complexity between them using equation (3).

Step 2. For each class, calculate its testing cost and test profit and set its test priority.

Step 3. Select appropriate classes according to their test priority and add them into the class test order.

Step 4. If the test order is incomplete, execute steps 2 to 3 repeatedly.

Step 5. Generate the completed class test order.

# 4 EVALUATION

## 4.1 Subjects Description

To evaluate the performance of our proposed CITO-generation approach ConCITO, we selected 10 open-source programs from SIR (Software-artifact Infrastructure Repository, http://sir.unl.edu/portal/index.html) as shown in Table 2, in which columns 1 through 6 show the program name, a description, the number of classes, the number of dependencies, the number of cycles, and the number of lines of code, respectively.

These programs are chosen from various domains of the software industry and have diverse functions. The number of classes between the various programs varies from tens to hundreds, and the number of direct dependencies varies from 36 for *daisy* to 1292 for *xstream__spl*. The sizes of these subjects in terms of LOC vary from 1148 for a simple program like *daisy* to over 32,882 for the complicated program *jmeter*. Because of the different sizes, these programs are highly representative as subjects for the CITO problem.

## 4.2 Implementation

To acquire the coupling information, we adopt Soot ⟨http://www.sable.mcgill.ca/soot/⟩, a Java program analysis framework, to identify and analyze the dependencies among the classes from the bytecode files. To collect information on interclass attribute coupling, we construct a control flow graph for each method and implement the interface *soot.toolkits.scalar.FlowSet* to save the data in the node of the control flow graph by overwriting some methods in the interface, such as *clone, clear, copy, isEmpty, intersection, difference,* and *union*. The data flow information can be collected by performing intraprocedural data flow analysis, and information on the interclass attribute coupling, such as the number of attributes accessed, can be obtained.

To collect information on interclass method coupling, we first identify the type to which the method invocation belongs: *static invocation, interface invocation, special invocation,* and *virtual invocation*. We then generate the method call graph according to its category. For each method call, we record its line number, store the invoked methods, and count how many times that they are invoked. After obtaining the information of interclass coupling, we can determine the interclass transitive relationship, construct the EORD, and extract the transitive relationship chains. For the length of transitive relationship chains, we set it as three, four and five, respectively, and compare statistics about different transitive relationship chains, such as their number and execution probability. The statistics show that transitive relationship chains that contain 'three' methods or attributes is the most common and important in practice. The more details are introduced in the sections 4.4. We performed all experiments on a Dell server with 32GB of memory and two 3.07GHz XEON X5675 CPUs using JDK 1.7.

## 4.3 Research Questions

Our experiments are designed to address the following five research questions:

**RQ1: Does the transitive relationship exist in practice?**

Iterative method invocation among classes might exist in the source code of the program from coding experience, but we cannot determine whether this kind of transitive relationship exists in the actual programs due to a lack of relevant research work in the field of CITO problems.

To answer this question, we first identify the interclass transitive relationship by extracting all transitive relationship chains of length three. We then collect information on transitive relationships for 10 subject programs to confirm the existence of the transitive relationship.

**RQ2: What is the most appropriate setting for the length of transitive relationship chains?**

Since the length of transitive relationship chains is an important factor for transitive relationship, we set parameter of length as three, four and five, respectively, and conduct statistics to make an appropriate setting for the following experiment.

**RQ3: Can the transitive relationship affect the generated class integration test orders?**

We adopt the CITO algorithm (Section 3.4) to generate CITOs for each program, considering transitive relationships and only considering direct relationships separately, and compare the two results to examine the effects of the transitive relationship on the generated CITOs.

TABLE 2
DESCRIPTION OF SUBJECT PROGRAMS

| Programs | Source | Description | Classes | Deps | Cycles | LOC |
|---|---|---|---|---|---|---|
| Ant | http://sir.unl.edu/content/bios/ant.php ant (v0) | Deployment tool | 200 | 720 | 3430 | 17,065 |
| Daisy | http://sir.unl.edu/content/bios/daisy.php daisy (v1.1) | NFS UNIX-like filesystem | 23 | 36 | 4 | 1148 |
| email__spl | http://www.infosun.fim.uni-passau.de/spl/hybrid/email__spl (original) | Email tool | 39 | 61 | 38 | 2276 |
| JaConTeBe | http://sir.unl.edu/content/bios/JaConTeBe.php JaConTeBe (v1.0) | Java Concurrency Testing Benchmark | 151 | 156 | 53 | 6550 |
| Jboss | http://sir.unl.edu/content/bios/jboss.php package used: org.jboss.management.mejb and j2ee of jboss (v0) | Application server | 91 | 83 | 1 | 5252 |
| Jmeter | http://sir.unl.edu/content/bios/jmeter.php jmeter (v0) | Load test tool | 372 | 1252 | 729 | 32,882 |
| log4j3 | http://sir.unl.edu/content/bios/log4j3.php log4j3 (v0) | Log tool for java | 261 | 784 | 2173 | 15,665 |
| notepad__spl | http://sir.unl.edu/php/showfiles.php notepad__spl (v1.0) | Source code editor | 65 | 141 | 227 | 2419 |
| xml-security | http://sir.unl.edu/content/bios/xml-security.php xml-security (v0) | XML encryption | 239 | 759 | 98,236 | 18,702 |
| xstream__spl | http://xstream.codehaus.org/ xstream__spl (v0) | Serialize objects to XML | 414 | 1292 | 88 | 18,327 |

**RQ4: Can the transitive relationship affect the stubbing cost of the generated class integration test orders?**

If we confirm the influence of the transitive relationship on the generated CITOs, the effect of this kind of indirect relationship on the stubbing cost of test orders requires further study. To answer this question, for each program, we compare the stubbing cost of the two kinds of CITOs obtained in RQ3 in terms of the overall stubbing complexity, attribute complexity, method complexity, control complexity, and stubs and then analyze the effect of the transitive relationship on the stubbing cost according to the changes in the five quality indicators.

**RQ5: What is the performance of ConCITO to generate class integration test orders?**

We compare ConCITO with two methods: CITO-CM and the random iterative algorithm (RIA). CITO-CM, proposed by Jiang et al. [10], is a representative graph-based method that improved on the classical Briand's method and uses coupling measures to generate CITOs. The RIA, proposed by Wang et al. [11], can achieve satisfactory performance with a simulated annealing algorithm. We carefully implement these two methods and improve them: for CITO-CM, we assign the weight for each edge in the diagram based on the stubbing complexity that has considered the measurement of control coupling; and for RIA, we choose the improved formula of stubbing complexity that has considered the measurement of control coupling as the fitness function to direct the evolution for populations.

We use the overall stubbing complexity ($OCplx$) as the indicator to evaluate the effectiveness. $OCplx$ can be calculated with equation (4), and the test order with the minimum value of $OCplx$ is desired. To evaluate the efficiency, we measure the execution time for each method by the ratio of runtime ($RT$), which can be calculated with equation (5):

$$RT(x) = \frac{RunTime(x)}{RunTime(\text{ConCITO})} \quad (5)$$

where the runtime includes the time for static analysis and for generation of CITOs. To evaluate the efficiency, we measure the execution time for each method by the ratio of CITO generation. An RT value greater than 1 indicates that the time complexity of method x is higher; otherwise, ConCITO requires more time.

### 4.4 Experimental Results

In this section, we answer the five proposed research questions through analyzing the experimental results.

**RQ1: Does the transitive relationship exist in practice?**

We execute static analysis on 10 subject programs and identify the direct relationships and transitive relationships. Specifically, we identify transitive relationships by extracting all transitive relationship chains of length three. Fig. 6 shows the information of relationships for 10 subject programs. For each subject, the value above the bar represents the percentage of the transitive relationships in all relationships, including both transitive and direct relationships. Transitive relationship exists in all 10 subjects, and in seven of the 10 subjects, the percentage of the transitive relationships exceeds 40%. The highest proportion of transitive relationships (in *email__spl*) reaches 48.31%, and even the lowest proportion (*jboss*) is still 7.78%.

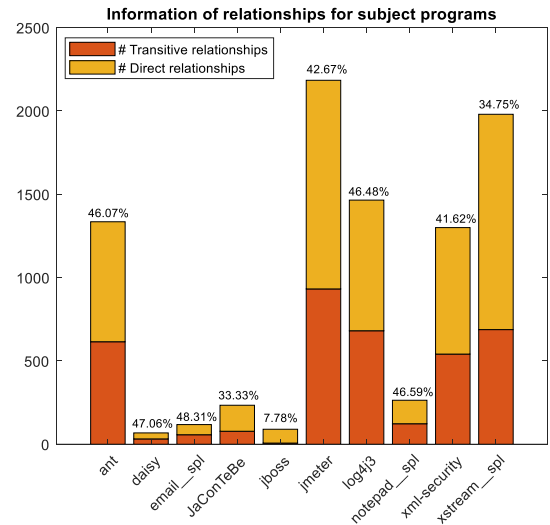

Fig. 6. Information of relationships for subject programs

To analyze the distribution of transitive relationships, we count the number of classes and methods involved in transitive relationships and not involved in transitive relationships, respectively. Fig. 7 shows the information for classes that are related to transitive relationships, and Fig. 8 shows the information for methods that are related to transitive relationships. As in Fig. 6, the value above the bar in Fig. 7 represents the percentage of the classes that are involved in transitive relationships in all classes, and the value in Fig. 8 indicates the proportion for which the methods involved in transitive relationships account. These two figures show that nearly 70% of the classes and more than 40% of the methods are involved in transitive relationships in the subjects, except *JaConTeBe* and *jboss*. The number of transitive relationships shows a gradually rising trend with the expansion in the scale of systems; that is, the number of classes and methods are both increasing. In *JaConTeBe*, classes that relate to transitive relationships constitute more than 50%, and such methods account for 24.72%. In *jboss*, the proportion is 9.89% for classes and 3.55% for methods, which are the smallest figures. In general, transitive relationships are common in the object-oriented programs regardless of the system's size and complexity.

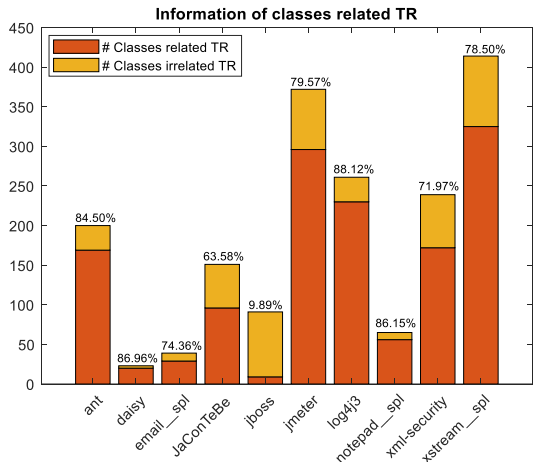

Fig. 7. Information of classes related TR

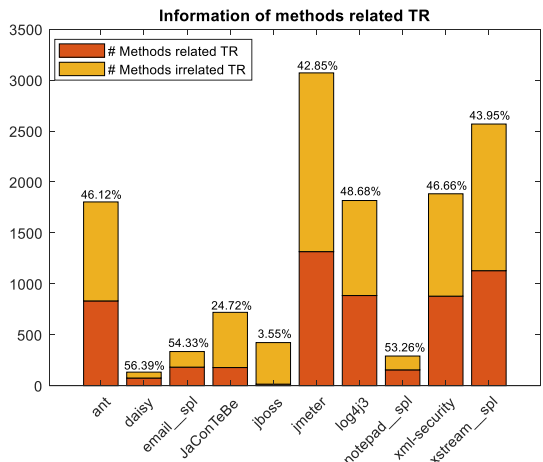

Fig. 8. Information of methods related TR

**RQ2: What is the most appropriate setting for the length of transitive relationship chains?**

Length of transitive relationship chains is an important factor to identify transitive relationships. We conduct statistics by setting the length of transitive relationship chains as three, four and five respectively, to study the impact of this setting on transitive relationships.

Table 3 shows the number of transitive relationship chains when their length is three, four, and five, respectively. We omit transitive relationship chains with two or more consecutive intraclass method calls or attribute assess, for example, class $A\rightarrow$class $A\rightarrow$class $A\rightarrow$class $C$, because this transitive relationship chain make less contributions to identify more transitive relationships. As shown in Table 3, the number of transitive relationship chains decreases with their length increases except for *daisy*. Although *daisy* is the simplest program with 23 classes and 36 class direct relationships among all programs, its transitive relationships are more complicated. Moreover, no transitive relationship chains with length more than three exist in *jboss*.

TABLE 3
NUMBER OF TRANSITIVE RELATIONSHIP CHAINS

| Programs | # TRC3 | # TRC4 | # TRC5 |
|---|---|---|---|
| ant | 2253 | 2002 | 1976 |
| daisy | 200 | 251 | 280 |
| email__spl | 423 | 419 | 313 |
| JaConTeBe | 133 | 34 | 9 |
| jboss | 13 | 0 | 0 |
| jmeter | 3818 | 2913 | 1875 |
| log4j3 | 1808 | 1162 | 696 |
| notepad__spl | 210 | 141 | 60 |
| xml-security | 1948 | 1584 | 1273 |
| xstream__spl | 1768 | 835 | 426 |

Fig. 9 and Fig. 10 show the number of transitive relationships for 10 programs, where blue bars represent the number of transitive relationships identified by transitive relationship chains of length three, orange bars represent the added transitive relationships by transitive relationship chains of length four, and yellow bars represent the number of transitive relationships increased by transitive relationship chains of length five on the basis of length four. Although transitive relationship chains of length four and five increase the number of identified relationships, transitive relationship chains of length three already identify a large part of transitive relationships, with the lowest proportion as 66.63% (681/ (681+201+140)) for *log4j3* and the highest proportion 100% for *jboss*.

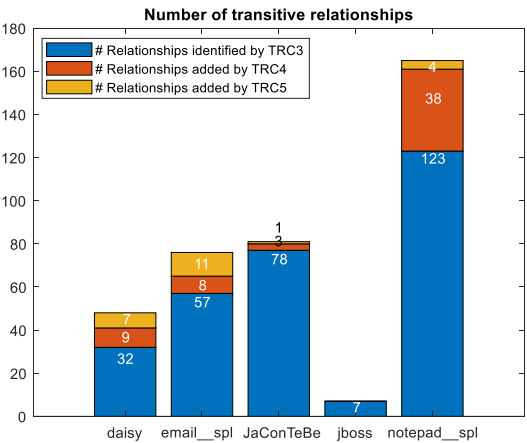

Fig. 9. Number of transitive relationships for five small programs

 

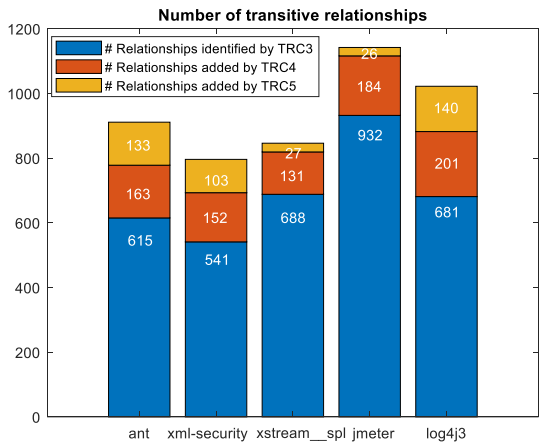

Fig. 10. Number of transitive relationships for five large programs

For all transitive relationship chains of length three, four and five, we calculate their execution probability as equation (2). Fig. 11 shows the distribution of execution probability of different sizes transitive relationship chains in nine programs (except for *jboss*). In terms of execution probability, we pay more attention to its proportion rather than its number. As shown in Fig 11, 27.47% of transitive relationship chains of length three are definitely executed, while for transitive relationship chains of length four and five, these figures are only 16.09% and 9.03%, respectively. What's more, the execution probability of 47.20% (27.47% + 1.56% +18.17%) transitive relationship chains of length three is above 0.5, while the proportion decreases to 29.20% (16.09%+1.23%+11.88%) and 19.95% (9.03%+1.72%+9.19%) for transitive relationship chains of length four and five. Consequently, more transitive relationship chains of length four and five are less likely to be executed compared with transitive relationship chains of length three. For example, the percentage of transitive relationship chains whose execution probability is lower than 1% is 20.40%, 32.61%, and 43.23% for length three, four and five, respectively. These statistics show that the more methods or attributes involved in the transitive relationship chain, the less likely that the program can be executed in accordance with the transitive relationship chain.

Overall, we have three findings from statistics. (1) For most programs, the number of transitive relationship chains of length three is the most. As the length rises, the number of transitive relationship chains decreases. (2) Transitive relationship chains of length three are able to identify more than 66.63% transitive relationship, especially, for four programs (*JaConTeBe*, *jboss*, *jmeter* and *xstream__spl*), this proportion exceeds 80%. (3) The proportion of transitive relationship chains that are more likely to be executed is much higher for length of three compared with length of four and five. Moreover, considering that no transitive relationship chains of length four or above in *jboss*, transitive relationship chains that contain 'three' methods or attributes is the most common and important in practice. Therefore, we set the length as three in our following experiment.

**RQ3: Can the transitive relationship affect the generated class integration test orders?**

We apply the CITO algorithm (Section 3.4) to generate CITOs for each subject in the context of considering two kinds of relationships and only considering a direct relationship, respectively. Our comparison and analysis of the differences between the results obtained in both situations is shown in Fig. 12 and Table 4.

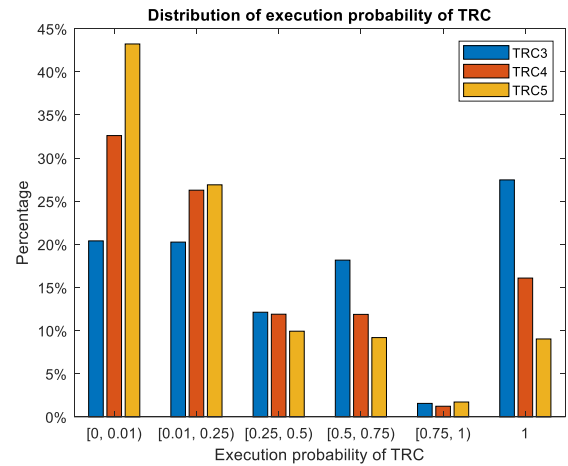

Fig. 11. Distribution of execution probability of TRC

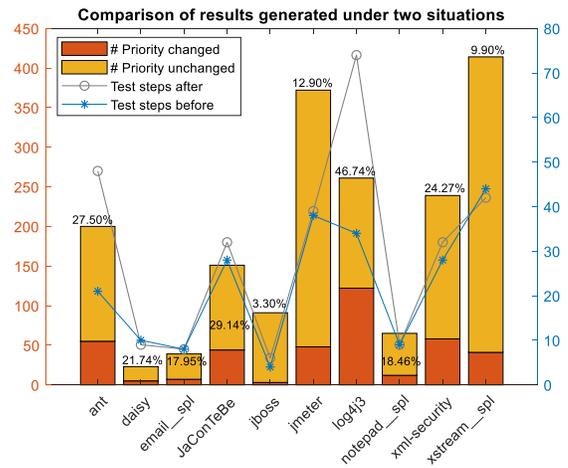

Fig. 12. Comparison of results generated under two situations

Fig. 12 counts the number of classes whose test priority changed in the test orders generated in the two situations and the number of steps for the two test orders. On the left vertical axis, the number of classes whose test priorities differ in the two test orders is represented by the orange bar, and the number of those whose test priorities were unchanged is represented by the yellow bar. The number of steps required for testing before and after considering a transitive relationship is on the right vertical axis. The percentage above the bar represents the proportion of classes whose test priority changed accounting for all classes. It is worth noting that the changes in test priority are relative. For example, class $C$ was integrated before class $A$ and $B$ in the test order generated considering only a direct relationship, that is, the test order of the three classes is $C \rightarrow A \rightarrow B$. However, in the latter test order, class $C$ was integrated after class $A$ and $B$. Although the relative orders of classes $A$ and $B$ have not changed, the test priority of the three classes differs from that in the former test order. To simplify the comparison, we only recorded the changes of test priority like class $C$. As shown in the bar chart, changes of test priority exist in all subjects; the test priority of more than 10% of the classes differs in two test orders except *jboss* and *xstream__spl*; especially in five subjects, *ant*, *daisy*, *JaConTeBe*, *log4j3*, and *xml-security*, the proportion of classes whose test priority changed exceeds 20%. The test steps of two test orders differ, except *email__spl* and *notepad__spl*.

TABLE 4
EXAMPLES OF PRIORITY CHANGES

| Programs | Examples | Programs | Examples |
|---|---|---|---|
| ant | 2 ant.Project (1→7)<br>95 ant.types.CommandlineJava (8→14)<br>154 ant.taskdefs.optional.junit.BatchTest (13→44) | jmeter | 33 jmeter.gui.util.MenuFactory (20→18)<br>153 jmeter.engine.StandardJMeterEngine(20→26)<br>194 jmeter.gui.action.Load (23→29) |
| daisy | 3 daisy.DaisyLock (2→3)<br>6 daisy.Petal (1→2)<br>15 DaisyUserThread (9→8) | log4j3 | 5 log4j.Category (4→14)<br>200 log4j.lf5.LF5Appender (8→61)<br>260 log4j.xml.XMLLayout (3→56) |
| email__spl | 5 TestSpecifications.SpecificationManager (7→5)<br>18 runspl.RunSPL (6→7)<br>23 featuremodel.Configuration (4→5) | notepad__spl | 20 smashed.Actions (2→4)<br>39 tests.NotepadTEST (7→3)<br>55 tests.NotepadTEST$1 (2→8) |
| JaConTeBe | 28 Derby5560 (12→8)<br>46 Groovy4736 (6→1)<br>67 Test120 (1→16) | xml-security | 10 xml.security.utils.XMLUtils (3→1)<br>72 xml.security.utils.Base64 (17→12)<br>224 xml.security.utils.SignerOuputStream(15→18) |
| jboss | 21 j2ee.StateManagement (1→2)<br>87 mejb.ManagementBean$LocalConnector(3→6)<br>88 mejb.ManagementBean (2→5) | xstream__spl | 30 xstream.core.JVM (27→7)<br>375 xstream.io.xml.TraxSource (9→12)<br>391 xstream.mapper.AnnotationMapper (19→14) |

TABLE 5
COMPARISON OF STUBBING COST

| Programs | CITO-DR | | | | CITO-TR | | | | |
|---|---|---|---|---|---|---|---|---|---|
| | OCplx | ACplx | MCplx | Stubs | OCplx | ACplx | MCplx | TCplx | Stubs |
| ant | 3.05 | 45 | 130 | 35 | **2.23** | 41 | 98 | 43.51 | 30 |
| daisy | 0.35 | 0 | 11 | 5 | **0.14** | 0 | 5 | 1.25 | 4 |
| email__spl | 0.89 | 4 | 44 | 10 | **0.77** | 4 | 44 | 51.13 | 12 |
| JaConTeBe | **2.53** | 4 | 86 | 44 | 5.77 | 10 | 88 | 28.91 | 44 |
| jboss | 0.94 | 2 | 3 | 3 | **0.13** | 0 | 2 | 0.0 | 1 |
| jmeter | 3.56 | 32 | 113 | 50 | **2.97** | 32 | 114 | 65.71 | 55 |
| log4j3 | **4.31** | 30 | 199 | 110 | 6.42 | 26 | 166 | 107.41 | 115 |
| notepad__spl | **1.66** | 6 | 67 | 45 | 2.30 | 5 | 65 | 32.13 | 49 |
| xml-security | 4.61 | 29 | 178 | 73 | **4.50** | 27 | 152 | 130.31 | 93 |
| xstream__spl | 2.71 | 4 | 126 | 75 | **1.93** | 2 | 89 | 59.56 | 73 |

Table 4 illustrates some examples of classes whose test priority differs in the two test orders and gives the information of the class number, the class name, and the change in its test priority. For instance, the test priority of the class *daisy.DaisyLock*, whose class number is 3 in the *daisy* system, was demoted from number 2 in the test order generated in the case that only considered a direct relationship to number 3 in the test order that considered a transitive relationship (third row, second column, first line).

From this discussion, we can determine that the test orders generated in the case that considered only a direct relationship and in the case that considered a transitive relationship are disparate. To confirm that these differences between two test orders are indeed caused by transitive relationships, we identify the direct relationships and transitive relationships that relate to classes whose test priority differs and collect and analyze the information on attribute coupling, method coupling, and control coupling in these relationships; we finally find that the transitive relationship can really affect the generated CITOs.

**RQ4: Can the transitive relationship affect the stubbing cost of the generated class integration test orders?**

According to the test orders generated in RQ3, we compare the stubbing cost in terms of overall stubbing complexity (OCplx; the sum of the complexity of stubs), attribute complexity (ACplx; the total number of emulated attributes), method complexity (MCplx; the total number of emulated methods), control complexity (TCplx; the sum of the control complexity of each stub), and stubs (the number of required stubs). Table 5 compares two test orders in the five quality indicators. The OCplx of CITOs generated by considering a transitive relationship (CITO-TR) is lower than that of CITOs generated by considering only a direct relationship (CITO-DR) for all subjects except *JaConTeBe*, *log4j3*, and *notepad__spl*.

Overall, the strategy performs better in most subjects but sometimes performs worse, as shown in Table 5, because multiple classes would have the maximum net test benefit during the calculation process by the CITO-TR approach after considering the transfer dependency. At this time, the class with the lowest test cost is selected as the current test class, which means that the final result cannot become the optimal solution.

In terms of ACplx and MCplx, the results in the two situ-ations are the same for *email__spl*. Except that CITO-TR emulated more attributes and methods for *JaConTeBe*, and the MCplx of CITO-TR is a bit higher for *jmeter*, the results of CITO-TR for other subjects are lower than those of CITO-DR. In systems, the more frequently a class accesses attributes or invokes methods in other classes, the more likely it indirectly accesses attributes or calls methods in a third class, and thus the more likely it would form transitive relationships with others. Our proposed formula for stubbing complexity not only considers a transitive relationship, it also emphasizes the complexity of attribute and method when measuring the stubbing cost for relationships that involve a large number of attribute coupling and method coupling. Therefore, this kind of relationship would not take precedence in CITO-TR due to its high stubbing complexity, and CITO-TR gives better results.

The value of TCplx is not zero for all subjects except *jboss*, which indicates that the transitive relationship indeed affects the stubbing cost of CITOs. The attribute complexity and method complexity of some stubs are zero, but the control complexity is not, so the number of stubs of CITO-TR is greater for some programs than those of CITO-DR.

To further explain the influence of the transitive relationship on the stubbing cost of the generated CITOs, we select three programs, *JaConTeBe*, *log4j3*, and *notepad__spl* to compare the stubs constructed according to the two test orders and attempt to determine why the overall stubbing complexity of CITO-TR is higher than that of CITO-DR.

Excluding the same stubs, the differences in the stubs created by CITO-DR and CITO-TR for three programs are shown in Tables 6 to 8, where columns 1 and 2 represent the information of stubs created by CITO-DR and CITO-TR, respectively. Column 3 describes the changes in the stubs created by CITO-TR compared to the stubs created by CITO-DR. In each table, the value to the left of the arrow is the source class number, and the value to the right of the arrow is the target class number that depends on the source class. The values in parentheses indicate the ACplx, MCplx, and TCplx of the stub, respectively. For instance, according to the class test order generated for *JaConTeBe* by CITO-DR in Table 6, a stub should be created for class 136 to emulate the behaviors of class 135; the values of ACplx, MCplx, and TCplx are 0, 1, and 2.0 (second row, first column, fourth line). Although the control complexity of the interclass transitive relationship is not estimated in CITO-DR, for easy comparison, TCplx of stubs created by CITO-DR are calculated according to the method mentioned in Section 3.2.

TABLE 6
COMPARISON OF STUBS FOR JACONTEBE

| Stubs created by CITO-DR | Stubs created by CITO-TR | Differences |
|---|---|---|
| 31-->30--(0,1,1.5)<br>31-->32--(0,1,1.5)<br>67-->68--(0,3,4.0)<br>136-->135--(0,1,2.0) | 30-->31--(0,3,0.0)<br>32-->31--(0,3,0.0)<br>68-->67--(4,0,0.0)<br>135-->136--(2,2,0.0) | ACplx:+6<br>MCplx:+2<br>TCplx:-9 |

For *JaConTeBe*, Table 6 shows that our method constructed the same number of stubs under two situations and that the classes involved in the stubs are basically identical. To reduce the control complexity, CITO-TR constructed completely opposite stubs, and as a result, the values for attribute complexity and method complexity increased by six and two, respectively, compared with CITO-DR.

Table 7 shows that for *log4j3*, the values of attribute complexity, method complexity, and control complexity of the test order generated by CITO-TR decreased by 4, 33, and 25.5, respectively, compared with those generated by CITO-DR. However, due to the newly constructed five stubs, the overall stubbing complexity of CITO-TR is a bit higher than that of CITO-DR.

For *notepad__spl*, Table 8 indicates that the stubs created by CITO-TR emulate one fewer attribute and two fewer methods than the stubs of CITO-DR, but because the value of control complexity increased by 5.6, its overall stubbing complexity is higher than the latter because multiple classes have the maximum net test benefit during the calculation process with the CITO-TR approach after considering transfer dependency. At this time, the class with the lowest test cost is selected as the current test class, which means that the final result cannot become the optimal solution.

The above analysis shows that the overall stubbing complexity can measure the stubbing cost of various CITOs, but hides the details of stubs. Thus, we should resort to the other four factors (i.e., ACplx, MCplx, TCplx, and stubs) to compare the performances. In most cases, these four factors restrict each other: while reducing the value of one factor, the increase in another factor cannot be avoided, as previously shown. In practice, testers can select the appropriate class integration test orders based on the existing test conditions to satisfy their practical demands and meanwhile achieve the aim of reducing the testing efforts.

TABLE 7
COMPARISON OF STUBS FOR LOG4J3

| Stubs created by CITO-DR | Stubs created by CITO-TR | Differences |
|---|---|---|
| 5-->12--(5,20,12.8)<br>213-->27--(0,3,0.0) | 5-->35--(0,0,1.0)<br>213-->13--(0,0,0.3) | ACplx:-5<br>MCplx:-23<br>TCplx:-11.5 |
| 18-->12--(0,3,8.0)<br>32-->27--(0,3,0.0)<br>42-->35--(1,0,0.0)<br>231-->230--(1,2,2.5)<br>248-->247--(0,2,2.0) | 12-->18--(4,0,0.0)<br>27-->32--(0,0,0.1)<br>35-->42--(0,3,0.0)<br>230-->231--(2,1,0.0)<br>247-->248--(2,0,0.0) | ACplx:+6<br>MCplx:-6<br>TCplx: -12.4 |
| 116-->117--(2,14,6.8)<br>116-->123--(2, 2, 0.6)<br>116-->155--(0, 9, 7.5) | 116-->176--(0,0,2.0)<br>116-->177--(0,0,2.0)<br>116-->178--(0,0,2.0)<br>116-->186--(0,0,8.0) | ACplx: -4<br>MCplx:-25<br>TCplx:-8.1<br>Stubs:+1 |
| 2-->8--(3,2,0.0)<br>2-->9--(2,2,0.0)<br>2-->14--(0,1,0.0) | 27-->15--(0,3,0.1)<br>27-->16--(0,0,2.0)<br>27-->20--(0,4,0.0)<br>27-->29--(0,1,1.0)<br>57-->64--(0,5,0.6)<br>73-->81--(4,12,2.8)<br>73-->82--(0,1,0.0) | ACplx: -1<br>MCplx: +21<br>TCplx: +6.5<br>Stubs:+4 |
| Total | - | ACplx:-4<br>MCplx:-33<br>TCplx:-25.5<br>Stubs:+5 |

TABLE 8
COMPARISON OF STUBS FOR notepad__SPL

| Stubs created by CITO-DR | Stubs created by CITO-TR | Differences |
|---|---|---|
| 20-->18--(0,1,0.0) 20-->39--(0,2,0.0) 20-->57--(1,3,0.0) | 19-->57--(0,0,2.0) 19-->58--(0,0,2.0) 34-->22--(0,1,0.0) 34-->23--(0,1,0.0) 39-->8--(0,0,0.1) 39-->17--(0,1,0.5) 39-->55--(0,1,1.0) | ACplx:-1 MCplx:-2 TCplx:+5.6 Stubs:+4 |

From the results, we can conclude that the transitive relationships indeed have an effect on the stubbing cost of the generated CITOs; for most subjects, considering the transitive relationship when generating CITOs can reduce the stubbing cost to some extent.

**RQ5: What is the performance of ConCITO to generate class integration test orders?**

We compare ConCITO with CITO-CM and RIA in terms of the OCplx in equation (4) and the RT in equation (5). Each approach is executed 30 times for each system, and Table 9 shows the average of all 30 results. Note that, according to equations (4) and (3), our measure accounts for control coupling. Therefore, to avoid unfair comparisons, we take the effective complexity of the generated stubs as the indicator to compare the different approaches in this paper.

We conduct the statistical analysis for the 30 executions and use the boxplots of the 30 executions to report such data to evaluate the proposed approach more intuitively. Due to the limited space, we only present the total distribution of OCplx and RT for 10 projects using three approaches in Fig.13 and Fig.14, respectively. More details about the distribution of OCplx and RT for each project are upload with the raw data on GitHub (https://github.com/zzzyzz/ControoC-ITO).

TABLE 9
COMPARISON OF RESULTS WITH OTHER METHODS

| Programs | OCplx | | | RT | |
|---|---|---|---|---|---|
| | CITOCM | RIA | ConCITO | CITOCM | RIA |
| ant | **2.19** | 3.17 | 2.23 | 2.13 | 3.79 |
| daisy | 0.17 | 0.36 | **0.14** | 1.04 | 1.03 |
| email__spl | 0.86 | **0.48** | 0.77 | 1.18 | 1.34 |
| JaConTeBe | 8.62 | 6.02 | **5.77** | 1.82 | 6.58 |
| jboss | 0.58 | **0.13** | 0.13 | 1.01 | 3.23 |
| jmeter | 3.56 | 5.46 | **2.97** | 2.15 | 14.56 |
| log4j3 | **4.62** | 8.76 | 6.42 | 20.34 | 11.87 |
| notepad__spl | **2.15** | 2.17 | 2.30 | 7.37 | 1.31 |
| xml-security | 5.67 | 5.42 | **4.50** | 1765.15 | 4.76 |
| xstream__spl | 2.40 | 5.23 | **1.93** | 3.57 | 31.98 |

We can see that the OCplx of the test orders generated by ConCITO was the minimum in six of the 10 subject programs. In other four subjects, CITOCM can generate test orders with the least overall stubbing complexity for *ant*, *log4j3*, and *notepad__spl*; RIA outperformed other methods for *email__spl*; the performance of ConCITO was better than that of RIA for these programs except *notepad__spl*, ranking second.

The execution time of ConCITO was much less than that of other methods for all programs because our method does not include the process of cycle searching and breaking or the evolution of population. The unning time of CITOCM is closely related to the cycles, including the cycles' number and complexity. The programs *daisy* and *jboss* required less running time because they only contain six and one cycles, respectively, even considering the transitive relationship, whereas *xml-security* required the longest time due to its large number of cycles. The running time for RIA is intimately related to classes, including the number of classes and the interclass relationships. The execution times of *daisy*, *email__spl*, and *notepad__spl* were lower because the number of classes for the three systems was much less than 100; whereas *xstream__spl* was the slowest because it had the largest number of classes.

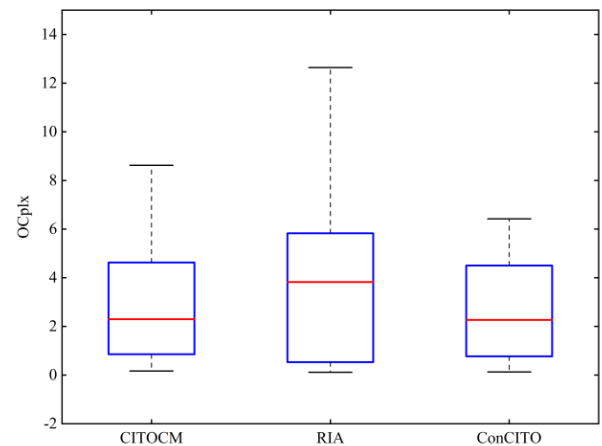

Fig. 13. Distribution of OCplx obtained by three approaches

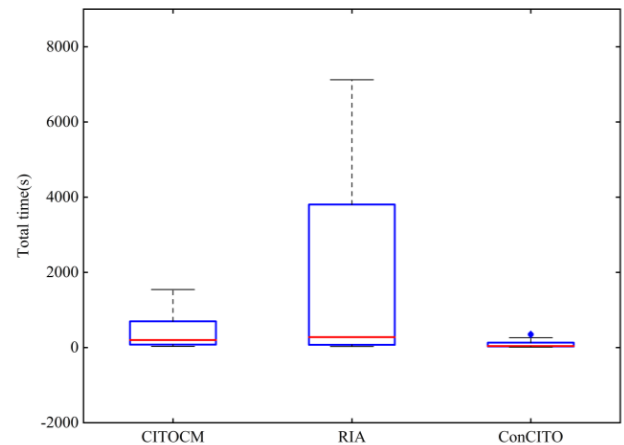

Fig. 14. Distribution of run time for three approaches

We apply the Wilcoxon signed-rank test [12] to evaluate whether significant differences exist between the three investigated approaches, for the OCplx and execution time. The Wilcoxon signed-rank test is a non-parametric test for two related variables that does not require the assumption of normal distributions. We make the null hypothesis $H_0$ that the distributions of OCplx and execution time for each approach are equal, whereas the alternative hypothesis $H_1$ is that the distributions are different. We set the significance level as 0.05. When a P-value is less than 0.05, we reject the null hypothesis and that is, there is a significant difference between the two approaches in terms of OCplx (execution time).

The P-values of OCplx and RT are shown in Table 10. CITOCM and ConCITO are deterministic for ten programs, and the P-value (0.203) show that there are no significant differences between the OCplx obtained by CITOCM and ConCITO, which is consistent with our results that ConCITO outperforms CITOCM on only seven programs. The P-value of OCplx is 0.038 for RIA, which means the OCplx obtained by RIA and ConCITO are statistically significant different. For RT, the P-values are 0.005 for both CITOCM and RIA, which means ConCITO and other approaches show significant differences in terms of execution time.

TABLE 10
P-VALUES OF OCPLX AND RT

| Approaches | OCplx | RT |
| --- | --- | --- |
| CITOCM | 0.203 | **0.005** |
| RIA | **0.038** | **0.005** |

From these results, we can conclude that ConCITO generates CITOs with the lowest stubbing cost in the shortest time for more than half of the programs; for most other subjects, the overall stubbing complexity of CITOs generated by ConCITO is lower. Therefore, ConCITO can obtain satisfactory results in a relatively short period of time in the situation that accounts for the transitive relationship.

## 4.5 Threats to Validity

Threats to internal validity arise when extracting path conditions and measuring the complexity of control coupling. ConCITO adopted the control dependence calculation algorithm [13] to identify the control dependences between nodes in the control flow graph, obtained the branch statements in nodes and extracted path conditions among them with Soot. In most cases, path conditions acquired with Soot are accurate; but some path conditions may be omitted due to lack of code standards and incompatibility of the source code with the Soot version. In response to this issue, for systems with fewer than 100 transitive relationships, we manually verified the source code and extracted path conditions for each call statement, and we did not find any missing path conditions; for other systems, we randomly selected 10% of transitive relationships and checked the path conditions for their relevant call statements, and we finally obtained the complete set of path conditions.

ConCITO measures the complexity of control coupling at a coarse-grained level. It calculates the execution possibility of call statements according to the probability that the relevant path conditions can be satisfied, but does not refer to the code context. For a call statement, an assignment statement in its execution path might declare the value of a variable before the call operation, so that the path condition of the call statement must be satisfied, and its execution possibility would be one. ConCITO does not examine the related source codes that might affect the truth or falsehood of the path conditions because our method aims only to propose a more comprehensive measurement for the stubbing cost.

A threat to external validity is that the experimental results cannot be generalized. To ensure that the experimental subjects are highly representative, we selected 10 subjects with various functions and sizes. The statistic test results show that no significant differences between the OCplx obtained by ConCITO and CITOCM, because they are both deterministic and all results are equal for the same program. Although our experimental subjects cannot cover all fields of software, the experimental results are still fairly versatile, and our conclusions could hold for all programs.

## 5 RELATED WORK

We focus our discussion on the most relevant work in three aspects: the generation of CITOs, current mocking practices, and the calculation of call statement execution probabilities.

### 5.1 Class Integration Test Order Generation

Works that address CITO problems consist of graph-based and search-based methods, which can also be characterized by their measurement of stubbing cost: the number of stubs, the stubbing complexity, and the stubbing cost measurement based on the Pareto model.

#### 5.1.1 Number of Stubs

Kung et al. [1] first used the number of stubs to assess CITOs and were the first to propose a graph-based method for CITO problems. They constructed an ORD and identified all cycles in the program. CITOs can be generated by topological sorting if no cycles exist in the diagram; otherwise, some edges must be deleted to break the cycles before that. Briand et al. [7] improved the approach used by Kung et al. and assigned weight for each edge to roughly estimate the number of cycles that it is involved. The edge with the highest weight is removed because it breaks the most cycles.

#### 5.1.2 Stubbing Complexity

Briand [4] first applied a genetic algorithm to generate a class test order with minimum stubbing complexity. Various kinds of evolutionary algorithms, such as the simulated annealing algorithm [14, 15] and particle swarm optimization [16], were then applied, and some performed better performance than the traditional ones. The random iterative algorithm [11] used as the comparison in our experiments was improved based on the existing simulated annealing algorithm. It exchanged positions of adjacent classes in the test orders to generate new individuals and accepted a better individual based on stubbing complexity. Compared with previous measurement of stubbing cost (i.e., number of stubs), these search-based approaches adopted a more refined estimation that can describe the emulated behaviors in stubs. However, search-based approaches may present a premature convergence problem and then trap into the local optimum. This issue does not arise with our method because each class in the test order is selected in turn according to its performance.

In graph-based methods, Hashim [17] estimated inter-class coupling strengths by the numbers of attributes and methods involved in interclass interaction. The fewer attributes or methods involved indicates a weak coupling strength between classes and thus a lower stubbing cost. Bansal [18] combined the two measurements of stubbing cost mentioned above (i.e., the stubbing complexity and the number of stubs). For each edge, he estimated the stubbing complexity and the number of cycles in which it is involved, deleting the edge with minimum stubbing complexity and involved in more cycles. Wang et al. [19] proposed a multigranular flow network model to describe inter-class relationships, and identified critical classes that are frequently executed, have more complexity, and are appreciably influential in the system. They evaluated each edge from three new aspects other than the stubbing complexity: vulnerabilities, threats, and failure consequences. The edge with the greatest weight was removed, and critical classes would be integrated first. The factors of stubbing complexity mentioned in the above graph-based methods, regardless of some coupling measures information, such as the number of parameters, or other elements that are used to estimate critical classes, only describe the direct interclass relationship, and measure the dependence between two classes. Our method not only estimates the coupling strength between two classes, it also considers the interclass relationship established via the third class.

### 5.1.3 Stubbing Cost Measurement Based on Pareto Model

Cabral et al. [20] first adopted the Pareto model to measure the stubbing cost. CITOs were estimated by the Pareto dominance concept and presented a good trade-off between distinct coupling measures (i.e., attribute complexity and method complexity). This measurement was commonly adopted in multi-objective optimization algorithms for CITO problems, such as NSGA-II, SPEA2 and PAES [21, 22], as well as hyper-heuristics. Guizzo et al. [23] proposed an online hyper-heuristic to provide multiple alternative choices of crossover and mutation for each evolution, and the best combination of two operations was selected to produce an appropriate approach to generate CITOs. Mariani et al. [24] advanced an offline hyper-heuristic based on grammatical evolution to automatically devise a multi-objective optimization algorithm for CITO problems; it obtained satisfactory results. Although stubbing cost measurement based on the Pareto model estimates attribute complexity and method complexity, respectively, it essentially assesses direct interclass relationship and omits transitive relationship proposed in our method.

### 5.2 Current Mocking Practices

This section shows the use of mocks in practice and further discusses the applicability of the stubbing complexity.

Spadini et al. [25] conducted an empirical study on mocking practices and found some interesting and valuable results regarding which dependencies testers mock or not and the corresponding reasons for their choices. Their findings are related to this work and provide some useful information for class integration test order generation to select its criteria.

Spadini et al. interviewed developers from three open-source software projects and one industrial system, and found that unlike unit test, the interviewees did not use mocks in integration testing. If testers mock a class, they are unable to ensure whether the real class actually works well with the existing integrated components. An ideal class test order requires no stubs or mocks, which is consistent with the opinion. However, cycles among class dependencies are common in a program as referred by Melton et al. [26]; for most cases, the class being tested depends on unreliable classes that have not been integrated. Stubs or mocks are created to provide services for the class being tested if the dependent classes are unavailable. Even if they are available, to make the test results reliable, the class being tested requires repeat verification when the dependent classes are integrated.

Therefore, a reasonable class integration test order is useful in practice to reduce time and efforts on repeat verification. Stubbing complexity may not be an appropriate term because mocks are not used in integration, but it is still useful since it reflects the intensity of class dependencies, which is the key point of integration testing to verify the interactions between classes. Interactions between classes are a main factor that affects the order in which classes are integrated regardless of whether we actually create mocks or stubs.

In addition, Spadini et al. also indicates that coupling among classes causes one of the main challenges in testing using mocks. Mocking can be difficult or even impossible when classes are trapped into complex coupling and abundant dependencies. Existing stubbing complexity measurement cannot describe the influence of class dependencies, but control coupling addresses this issue and helps to identify interclass indirect relationships. This opinion supports our consideration of control coupling in class integration test order generation.

### 5.3 Calculation of Call Statement Execution Probability

We remark that our method for setting the call statement execution probability is preliminary when the information from static analysis cannot be obtained, such as information about the range of variables in the path condition. Liu et al. [27] presented an approach to combine symbolic execution with volume computation to compute the exact execution probability of program paths and branches. Wang et al. [28] presented an approach to compute the probability of program paths. Luckow et al. [29] presented exact and approximate symbolic execution techniques for the probabilistic analysis of nondeterministic programs. Their methods are all complex. To keep the tradeoff between the complexity of probability paths and the precision of the analysis results, we choose use our method to compute the probability paths. Furthermore, the probability paths are not the main content of the paper and are thus left for a future study.

## 6 CONCLUSION AND FUTURE WORK

Because the existing measurement method for stubbing complexity evaluates mainly data coupling but omits control coupling, which is also common in interclass relationships, we introduced an integration test order strategy to consider control coupling called ConCITO. We defined the transitive relationship to describe the interclass indirect relationship, constructed an extended object relation diagram to depict interclass relationships more comprehensively, and measured the control coupling for a transitive relationship. Finally, a new measurement method was proposed to calculate the stubbing complexity that can be applied in algorithms of CITO generation. To evaluate ConCITO, we performed experiments on 10 programs and compared them with two approaches. The results show that control coupling altered the test priority of the class in the integration test orders and affected their stubbing cost; the new proposed approach, ConCITO, generated satisfactory class integration test orders in a relatively short time. In a future study, we will propose a fine-grained measurement for control coupling, which refers to the code context that is related to call statements and improve the formula for control complexity to make it more accurate and comprehensive.

## ACKNOWLEDGMENTS

The authors would like to thank the anonymous reviewers and editors for suggesting improvements and for their very helpful comments. This work is supported in part by the National Natural Science Foundation of China under grant No. 61673384, 61502497; and Natural Science Foundation of Jiangsu Province under grant No. BK20181353. Yanmei Zhang is the corresponding author. This work is also supported in part by the General Research Fund of the Research Grants Council of Hong Kong (No.11208017) and the research funds of City University of Hong Kong (9678149 and 7005028), and the Research Support Fund by Intel (9220097).

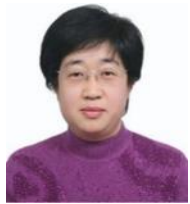

**Shujuan Jiang** was born in Laiyang City, Shandong Province, in 1966. She received the B.S. degree in Computer Science of Computer Science Department from East China Normal University, Shanghai, in 1990 and the M.S. degree in Computer Science from China University of Mining and Technology, Xuzhou, Jiangsu, in 2000. She received the Ph.D. degree in Computer Science from Southeast University, Nanjing, Jiangsu, in 2006.

From 1995 to today, she has been a Teaching Assistant, Lecturer, Associate Professor and Professor in the Computer Science Department, China University of Mining and Technology, Xuzhou, Jiangsu. She is the author of more than 80 articles. Her research interests include software engineering, program analysis and testing, software maintenance, etc. She is a Member of the IEEE.

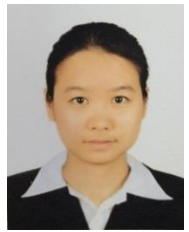

**Miao Zhang** is currently a PhD candidate under the supervision of Dr. Jacky Wai Keung in the Department of Computer Science, City University of Hong Kong. She received the M.S. degree in Computer Application Technology from the China University of Mining and Technology in 2018, and the B.S. degree in Software Engineering from Shandong University in 2015. Her research interests include software analysis, software integration testing and software maintenance. She is a Member of the IEEE.

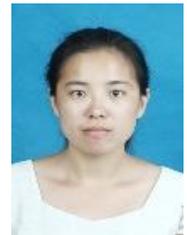

**Yanmei Zhang** received the B.S. degree in Computer Science of Computer Science Department from the Nanjing Tech University, in 2007 and the Ph.D. degree in Computer Science from China University of Mining and Technology, Xuzhou, Jiangsu, in 2012.

From 2007 to 2009, she was a M.S. candidate in Computer Science from China University of Mining and Technology, Xuzhou, Jiangsu. From 2012 to 2017, she has been a Lecturer, and since 2018, she has been an Associate Professor in the Computer Science Department, China University of Mining and Technology, Xuzhou, Jiangsu. She is the author of one book, more than 10 articles, and more than 2 inventions. Her research interests include software engineering, program analysis and testing, software quality, and intelligent evolutionary algorithm, etc.

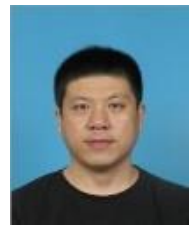

**Rongcun Wang** was born in Dezhou City, Shandong Province, in 1979. He received the Ph.D. degree in Computer Science and Technology from Huazhong University of Science and Technology in 2015. Since 2015, he has been a Lecturer in the Department of Computer Science, China University of Mining and Technology, Xuzhou, Jiangsu. He has published more than 10 articles and 3 inventions. His research interests include software testing, software maintenance, and software safety analysis.

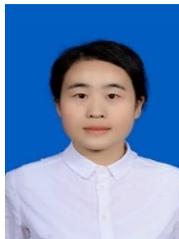

**Qiao Yu** was born in Laiyang City, Shandong Province, in 1989. She received the B.S. degree in School of Information from Shandong University of Science and Technology, Qingdao, Shandong, in 2012 and the Ph.D. degree in School of Computer Science and Technology from China University of Mining and Technology, Xuzhou, Jiangsu, in 2017.

From 2012 to 2014, she was a M.S. candidate in School of Computer Science and Technology from China University of Mining and Technology, Xuzhou, Jiangsu. Since 2017, she has been a Lecturer in School of Computer Science and Technology, Jiangsu Normal University. She is the author of about 8 articles. Her research interests include software engineering, software analysis and testing, and machine learning, etc.

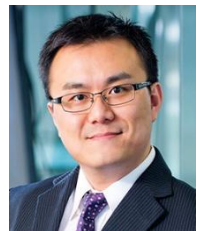

**Jacky Wai Keung** received the BSc (Hons) degree in Computer Science from the University of Sydney, and the PhD degree in Software Engineering from the University of New South Wales, Australia. He is Associate Professor in the Department of Computer Science, City University of Hong Kong. His main research interests include software effort and cost estimation, empirical modeling and evaluation of complex systems, and intensive data mining for software engineering datasets. He has published papers in prestigious journals including the IEEE Transactions on Software Engineering, the Empirical Software Engineering, and many other leading journals and conferences. He is a member of the IEEE.